\documentclass[sn-apa]{sn-jnl}

\jyear{2021}%

\usepackage{natbib}
\setcitestyle{numbers}
\usepackage{url}
\usepackage{hyperref}
\usepackage[T1]{fontenc}
\usepackage{apacite}

\usepackage[english]{babel}
\usepackage[utf8x]{inputenc}
\usepackage{graphicx}
\usepackage{wrapfig}
\usepackage{latexsym}

\usepackage{soul}
\usepackage[status=final,nomargin,inline,lang=spanish]{fixme}
\fxusetheme{colorsig}
\FXRegisterAuthor{dg}{adg}{Dario}
\FXRegisterAuthor{ac}{aac}{Atia}
\FXRegisterAuthor{sa}{asa}{Sergio}
\usepackage[export]{adjustbox}

\newcommand{\eg}{\emph{e.g.}, }       
\newcommand{\ie}{\emph{i.e.}, }      
\newcommand{\etal}{\emph{et al.} }         
\newcommand\etc{\emph{etc. }}

\begin{document}

\title[Signs for Ethical AI]{Signs for Ethical AI: A Route Towards Transparency}

 \author*[1]{\fnm{Dario} \sur{Garcia-Gasulla}}\email{dario.garcia@bsc.es}
 \author[1]{\fnm{Atia} \sur{Cort\'es}}\email{atia.cortes@bsc.es}
 \author[1]{\fnm{Sergio} \sur{Alvarez-Napagao}}\email{sergio.alvarez@bsc.es}
 \author[1,2]{\fnm{Ulises} \sur{Cort\'es}}\email{ia@cs.upc.edu}

\affil*[1]{\orgname{Barcelona Supercomputing Center (BSC)}, \orgaddress{\city{Barcelona},\country{Spain}}}

\affil[2]{\orgname{Universitat Politécnica de Catalunya - Barcelona-TECH}, \orgaddress{\city{Barcelona},\country{Spain}}}


\abstract{Today, Artificial Intelligence (AI) has a direct impact on the daily life of billions of people. Being applied to sectors like finance, health, security and advertisement, AI fuels some of the biggest companies and research institutions in the world. Its impact in the near future seems difficult to predict or bound. In contrast to all this power, society remains mostly ignorant of the capabilities and standard practices of AI today. To address this imbalance, improving current interactions between people and AI systems, we propose a transparency scheme to be implemented on any AI system open to the public. The scheme is based on two pillars: \textit{Data Privacy} and \textit{AI Transparency}. The first recognizes the relevance of data for AI, and is supported by GDPR. The second considers aspects of AI transparency currently unregulated: AI capabilities, purpose and source. We design this pillar based on ethical principles. For each of the two pillars, we define a three-level display. The first level is based on visual signs, inspired by traffic signs managing the interaction between people and cars, and designed for quick and universal interpretability. The second level uses factsheets, providing limited details. The last level provides access to all available information. After detailing and exemplifying the proposed transparency scheme, we define a set of principles for creating \textit{transparent by design} software, to be used during the integration of AI components on user-oriented services.}


\maketitle


\section{INTRODUCTION}

At the beginning of the XXI century, society is being forced to integrate a disrupting technology: Artificial Intelligence (AI). With striking resemblance, the same happened at the beginning of the XX century with another disrupting technology: Automobiles. Back in the 1900s, as cars were introduced to roads and streets, people had to integrate them in their everyday life. This caused significant conflict between the few who were profiting the most from the technology, and the rest \citep{eastman1971america}. The legislation had to be put in place to settle this conflict, a process that spanned several decades. 
Among the first regulations to be implemented was the registration of automobiles, which happened gradually between 1901 and 1920. It met with resistance from automobile associations. Driving licenses and speeding limits were also the subject of vivid debate until their benefits became evident to all agents. The most challenging of automobile regulations, the one which took the most to be implemented in a standardized manner, was the one responsible for organizing the interaction between people and automobiles in traffic. That is, traffic signs. 
A wide variety of traffic signs have been developed in the last century, and continue to change today, adapting to the new paradigms of mobility. The purpose of traffic signs is to limit the actions of automobiles to those safe to the public, but also to inform the public on how to interact responsibly with these powerful and potentially dangerous machines.


\begin{figure}[b]
\centerline{\includegraphics[height=3in]{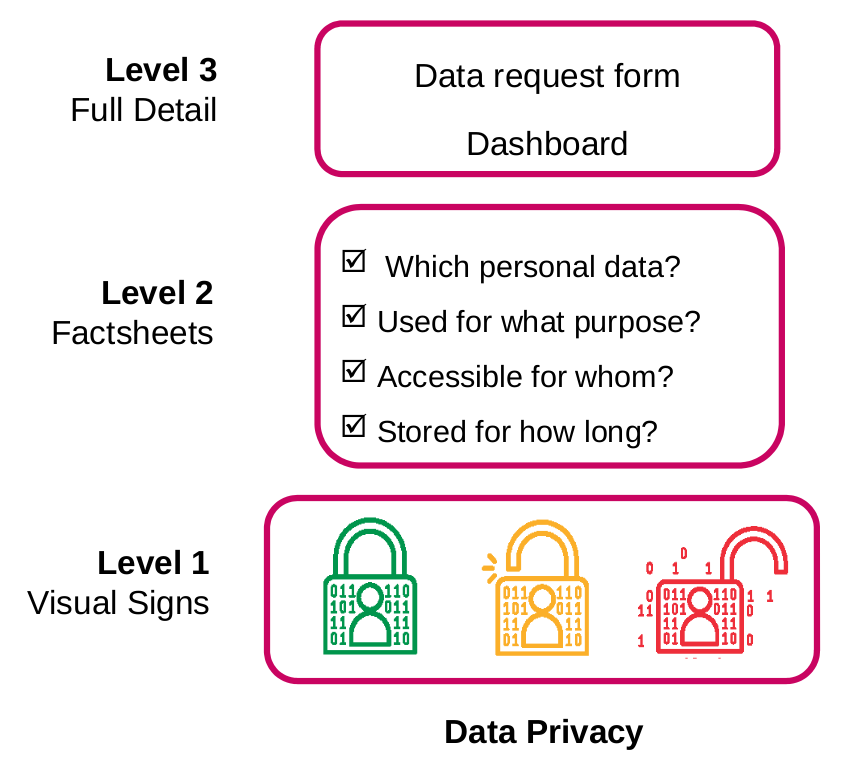}}
\caption{Summary of the scheme proposed for Data Privacy. Including three levels of information detail.} 
\label{fig:data_privacy}
\end{figure}

\begin{figure}
\centerline{\includegraphics[height=3in]{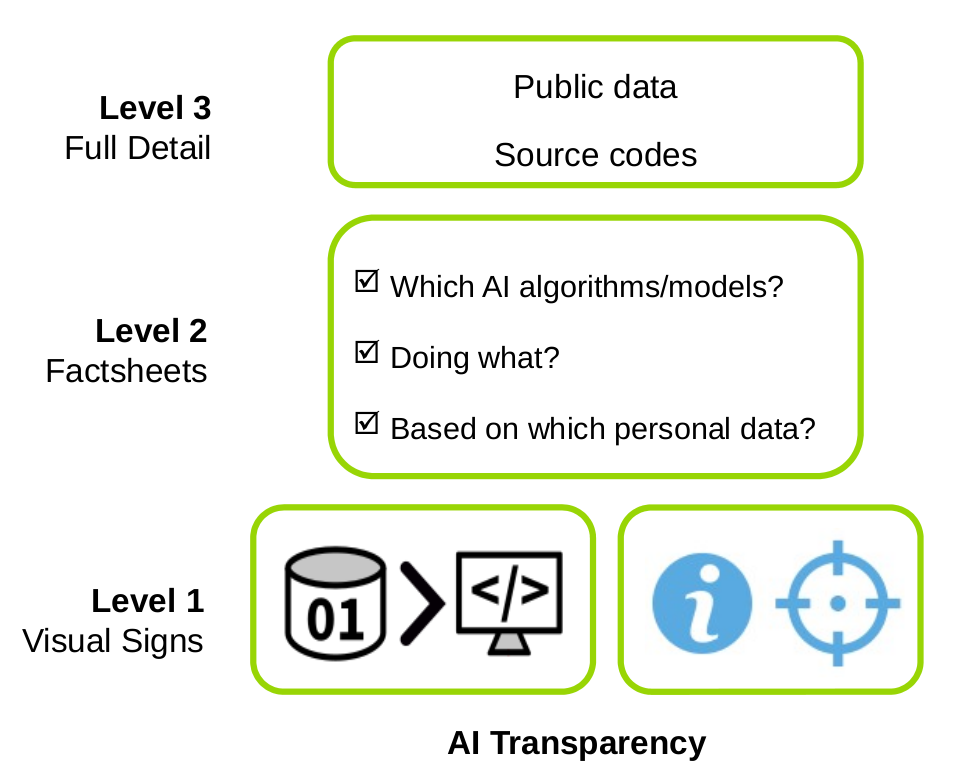}}
\caption{Summary of the scheme proposed for AI Transparency. Including three levels of information detail.} 
\label{fig:ai_trans}
\end{figure}

Today, society faces a similar challenge. AI is a disruptive technology that has started playing a significant role in our everyday life. A conflict of interests emerged between the few who are profiting from AI the most and the rest. Society is now in the process of creating legislation for AI. 
Looking ahead, if we consider the example from the previous century, it may take us decades until AI is correctly integrated into society. On the plus side, society may progress significantly by the end. One of the pending tasks on this journey is to define and implement proper means of interaction between the public and AI. We still need to find our \textit{traffic signs} for AI.

With this idea in mind, we start by looking at the nature and use of AI. At which aspects of AI should be transparent by fundamental ethical reasons. After a bibliography review described in Section \ref{sec:ethical}, we decided to split the problem between \textit{Data Privacy} and \textit{AI Transparency}. Data privacy considers the ethical and legal answers to questions like, who is gathering my data? For what purpose? AI Transparency considers only the ethical answers (since no legislation exists yet) to questions like, is there an AI algorithm running? What is it doing? Am I being targeted? 

From these two pillars, we design a complete transparency scheme, described in Section \ref{sec:scheme}, which covers each topic at three levels of depth. The first level, based on visual signs, is meant to be always visible, and accessible to everyone. The second level provides a degree of personalisation. 
Moreover, the third offers all the technical details available. A summary of the display is shown in Figures \ref{fig:data_privacy} and \ref{fig:ai_trans}. After the main contribution, Section \ref{sec:practice} exemplifies how it should operate in practice. Section \ref{sec:integration} introduces the \textit{transparency by design} principles, which can be used as blueprint for the development of AI systems. Then, Section \ref{sec:previous} reviews some works in a similar direction found in the bibliography. Finally, Section \ref{sec:conc} discusses the conclusions and future work.



\section{ETHICAL AND TRANSPARENT AI}\label{sec:ethical}

AI methods are not unethical by nature. Potentially unethical features, like those exploiting private information or producing biased responses, are inherent to AI methods required for services that our society values and demands. That being said, the use of AI can become unethical rather easily. The most straightforward way is to hide the existence of AI and its arguable properties from users. Thus, he first necessary step to enable ethical behaviour is \textit{transparency} since an ethical AI is only possible within a society aware of its existence. A society where users can make informed decisions on their interactions with AI systems, aware of what they are conceding and gaining.




The role of transparency within trustworthy AI is not straight-forward, as it depends on the context and does not follow a linear relation~\citep{felzmann2019transparency}. Turilli \etal identify two contradicting definitions of transparency~\citep{Turilli2009}, depending on the domain of application. From an ethical perspective, transparency relates to the notion of \textit{visibility} of the information (access, intended use or behaviour). At the same time, in the traditional IT domain, it also refers to the \textit{invisibility} towards the user. The purpose of the latter was to facilitate access from end-users who were not familiarised with the technology fueling computer systems. 
Its main objective was to make technology an appealing, user-friendly tool.  This approach to transparent technology has been satisfactory for many years, enabling the digitalization of society. However, with the generalized adoption of technology, and the emergence of AI systems, which demands more responsibility from users, the drawbacks of the invisibility approach have surpassed its benefits.

As of today, AI systems are often fed with personal or sensitive data, stored in data warehouses that are sometimes located on a third country or continent, and thus subjected to different regulations. This data is also often aggregated into new dataset that are used for other purposes unknown to the data owner (\ie the person itself). 
We are moving from an information society to an algorithmic society, where personal data is the new type of currency. In this new context, all end-users have an active role as data donors. Consciousness must be created regarding the potential impact of data transactions, both for the individual and the collective well-being. Otherwise, as it is already being disclosed on different scandals, we are bounded to a constant stream of data misuses, behavioural manipulations and massive discriminations. Unethical scenarios that ought to be opposed through the visibility approach to transparency.

\subsection{TRANSPARENCY IN ETHICAL AI LITERATURE}

The ethical relevance of transparency is evident in the literature. In \citep{Jobin19} authors review, 84 guidelines on ethical AI published around the world, by governments, high-tech giants or supra-national groups. Transparency, along with related terms such as explainability, interpretability, communication, disclosure and others, is the \textit{most} mentioned ethical principle, appearing in 73 out of 84 documents. In most of the reviewed literature, transparency is seen as a facilitator needed to achieve the rest of the principles of trustworthy AI systems. For example, if the information used to train an algorithm is public, it will contribute to reduce or avoid bias or unfair outcomes. In practice, transparency engages key ethical AI components such as accountability, traceability, justification, and proper assessment of capabilities and limitations of AI systems.

The first declaration of principles for the future research and development of AI was the Asilomar AI Principles \citep{Asilomar2017}. In these generic and high-level guidelines, transparency appears divided into two different ethical principles: failure transparency and judicial transparency. 
The former relates to the possibility to discover why did an AI system cause some harm, while the latter refers to explainability of any judicial decision-making process to an expert human. 

The definition and role of transparency have been further refined in more recent contributions. The High-Level Expert Group on Artificial Intelligence (AI HLEG) from the European Union has recently published a first draft of the Trustworthy AI Guidelines \citep{HLEG2019}, where transparency is one of its seven requirements. 
The guidelines propose a human-centred approach, to enhance end-users' confidence in using AI systems by offering means for a trustworthy interaction, respecting human agency and promoting governance mechanisms to ensure full accountability. The concept of transparency is here related to everything that is relevant to an AI system, \textit{i.e.} the data, the system and the business model. This includes providing means of \textit{(i)} traceability of the data and processes; \textit{(ii)} methods of explainability to understand both the technical aspects of the AI systems and the human decisions; and \textit{(iii)} channels of communication to inform users when interacting with an AI system. 

The IEEE has also published a draft version of their Ethical Aligned Design \citep{IEEE2019} principles and recommendations, being transparency one of the eight presented principles. Transparency is also linked to the level of disclosure of a decision. However, it focuses on decisions made by the system (or actions from a robot). In encompasses the concepts of explainability, traceability and interpretability.

Transparency is also one of the five principles of the OECD \citep{OECD2019} document of recommendations to build trustworthy AI systems. 
Transparency, along with explainability, is linked to a responsible disclosure of the information. The objective is to provide means to end-users and stakeholders to understand the outcomes of an AI system and make them aware of the interactions with these systems. 
In June 2019, the G20 Ministerial Meeting on Trade and Digital Economics published their declaration of principles and recommendations \citep{G202019}, based on the OECD document, and with a human-centred perspective.

Significantly, the guidelines here reviewed target a different set of stakeholders. Depending on whether the document is more or less human-centric, the responsibility is centred on the AI system or robot, such as is the case of Asilomar, or on the developers and users, such as in the HLEG case. At this point, it seems clear that all the stakeholders involved in the AI ecosystem must assume a certain degree of responsibility, and coherently, a certain level of transparency.
The management of organisations must assume transparency as a core value, including (and enforcing) it in their codes of conduct, for employees and members to follow. The designers and developers of AI systems need to integrate traceability and explainability as yet another necessary sanity check of their work. Last but not least, the end-users (or laypersons) need to be engaged in this process of learning how to interact with AI systems, understand the impact or origin of a decision and actively demand transparency processes. In general terms, it means to be responsible for the use of AI at every level of consumption.

The many existing stakeholders of ethical AI illustrates the need for a variable level of transparency, linked with the expertise of the user. For instance, a developer should be able to understand and explain the algorithms behind the AI system or justify the data set that has been used. 
However, should we demand a layperson to understand AI algorithms and data technologies to exercise their rights? Or is it enough to inform on the data processing and its outcomes? Such questions derive from the GDPR, and especially the Recital 71, which states that, under certain conditions, organisations must be able to explain to end-users the algorithms used to make decisions but does not specify which is the satisfactory level of explanation \citep{Selbst2017}, this is directly related with the \textit{right to an explanation}\citep{edwards2018enslaving}. 
From a technical perspective, while acknowledging recent advances~\citep{spagnuelo2019accomplishing,spagnuelo2020qualifying}, there is a generalized lack established metrics to assess the transparency of an AI system, an endemic limitation of the field that we try to at least bound in this work. 
Similarly, \citep{Wachter2016} raises the question about the appropriate timing to inform the user. Should it be before the automatic-decision process takes place, or after in order to explain the features or weights that were used to obtain a given outcome? Our stand in this regard is on the protectionist side, as discussed in Section \ref{sec:practice}.

A common aspect in all the existing guidelines and recommendations is that none of them are linked to a regulatory enforcement, or propose specific implementation mechanisms. To make the difficult transition from theory to practice, mechanisms need to be put in place to verify and certify AI systems. To advance in this direction, this paper makes a clear contribution through the complete specification of a transparency scheme for AI systems.


\subsection{Transparency in practice today}

The General Data Protection Regulation (GDPR\citep{GDPR2017}) was adopted in 2016 by the European Union and the European Economic Area. It is to be actively regulated since May 2018. GDPR regulates transparency and personal data and recognizes the new rights of citizens. These include the right of access (knowing which data of you has been stored) and the right to data erasure, also known as the right to be forgotten (having your personal
data deleted under certain circumstances). 
Since these considerations have been globally overseen so far, the implementation of GDPR in all its scope represents a daunting task.

As a reaction to GDPR, the Interactive Advertising Bureau (IAB) developed a unified implementation framework, intending to provide a standard for all ad industry. The IAB is an organization of over 600 companies directly or indirectly dedicated to online advertising and data gathering, mostly from Europe and the United States. 
It includes major technological companies, such as Amazon, Facebook, Google, IBM or Uber. The implementation proposed by the IAB is known as the Transparency and Consent Framework (TCF \citep{europe2019iab}), with version 1.0 being released on April 2018. After this was found insufficient to cover several aspects of GDPR \citep{ICO_report}, TCF 2.0 was released in August 2019. A public repository of TCF 2.0 is available\footnote{https://github.com/InteractiveAdvertisingBureau/GDPR-Transparency-and-Consent-Framework/blob/master/TCFv2/TCF-Implementation-Guidelines.md} for the industry to integrate with it. To the best of our knowledge, no official statement has been made yet regarding its GDPR compliance, but motivated concerns have been expressed\citep{matte2020purposes}.

The scientific community has analyzed the shortcomings of the current practice of user consent and transparency.  Recently, \citep{nouwens2020dark} gathered data from the five most popular providers of consent forms in the UK, and found that only 11.8\% met the GDPR requirements. 
The concerns regarding TCF were also voiced by the UK Information Commissioner’s Office (ICO) in June 2019. In its report \citep{ICO_report}, the ICO states that \textit{"privacy information provided often lacks clarity and does not give individuals an appropriate picture of what happens to their data.  
While we recognise that provision of this information in the online environment can be challenging, this does not mean that participants can ignore the requirements of PECR (‘clear and comprehensive information’) and the GDPR."}. 
The Privacy and Electronic Communications Regulations, or PECR, is a European Commission directive, approved in 2003, which complements GDPR. Among other things, it specifies that information provided regarding data collection means and data uses must be explained clearly to the public~\citep{pecr}. 


\textit{Clarity} is difficult to measure. Nonetheless, we consider that current implementations of GDPR through TCF 2.0 are \textit{clearly} obscure for three main reasons. First, language remains too technical and inaccessible, for example, built on the understanding of technical concepts like \textit{cookies}. Communication at its simplest must not and cannot include technical terminology. Secondly, the detail of current TCF implementations does not provide clear information to the spectrum of public demand. 
A hierarchy of information needs to be established to serve all, from the least to the most meticulous users. Third and last, the integration of the consent form into user displays is ineffective, and frequently, counterproductive. Forms today are mostly based on pop-ups, intrusive and perturbing, which promotes their immediate dismissal. Information regarding transparency should be naturally integrated into interfaces, being minimally intrusive and always visible. All these flaws make current implementation ineffective in practice \citep{acquisti2005privacy,nouwens2020dark}. 


So far, we have discussed the state of data transparency in practice, but we have not introduced the context of AI transparency. Unfortunately, beyond what has been regulated by GDPR, AI transparency in practice is non-existent. 
A few organization promote it, like the AI Transparency Institute, but has no practical adoption. 
As of today, people are unable to know the algorithms and data fueling the AI systems that interact with them. Moreover, what is more, people are unable to know of the mere existence of AI systems interacting with them. To fix this, in this paper, we propose an AI transparency scheme. However, before that, let us clarify some terminology.


\section{TERMINOLOGY}\label{sec:terms}

Throughout the specification of the proposed transparency scheme, in the following section, we frequently refer to the concepts of system, AI service and purpose. Given the importance of these terms for the contextualization of the proposal, we detail the interpretation used in this work.

In the context of this paper, a \textit{system} is a computational entity with a graphical user interface that enable users to interact with it. Web pages and device applications are popular examples of a system under this interpretation. A system which contains one or more AI services is an AI system.

In contrast an, \textit{AI service} is a computational mechanism that uses AI technology while interacting with users and/or their data for a definite purpose. AI technology includes knowledge representation methods, machine learning, statistical learning, data mining, data analysis, analytics and other related fields. The consumer of the service can be a user, the system containing the service, or a third party. A service is an AI service even if AI only plays a secondary role in it. Frequently, different AI services performing different specific tasks are combined to provide high level functionalities. Nonetheless, this does not alter the granularity of the AI service definition. An AI service is defined by its purpose, not by how its used.

\subsection{Purpose}\label{subsec:purpose}

An effective definition of purpose is essential for implementing transparency and consent. Purpose states what will the user receive in exchange for its consent, making it the main driver behind the acceptance or rejection of access and use rights. 
The definition of purpose under GDPR was refined in \citep{purpose_lim}. Following this adopted opinion, the  purpose is to be specified without vagueness or ambiguity as to its meaning or intent, in such a way so as to be understood in the same way by everyone, including the private data owners. This means putting users at the centre of purpose definition. 

The industry still has to adopt this definition of purpose. This delay or resistance generates requests of enforcement from society \citep{complain}, which anticipates changes ahead. 
As of now, let us consider the purposes defined by TCF 2.0 \citep{europe2019iab} as a starting point. From a users' perspective, some of these purposes are excessively detailed, while others are insufficiently detailed. For practical reasons, we aggregate them as follows:

\begin{itemize}
    \item Access device storage and/or data 
\begin{itemize}
        \item TCF Purpose 1: Store and/or access information on a device
    \end{itemize}
    \item Ad selection or evaluation
\begin{itemize}
    \item TCF Purpose 2: Select basic ads
    \item TCF Purpose 3: Create a personalised ads profile
    \item TCF Purpose 4: Select personalised ads
    \item TCF Purpose 7: Measure ad performance
\end{itemize}
    \item Content selection, creation and/or evaluation 
\begin{itemize}
    \item TCF Purpose 5: Create a personalised content profile
    \item TCF Purpose 6: Select personalised content
    \item TCF Purpose 8: Measure content performance
\end{itemize}
    \item Market research
    \begin{itemize}
    \item TCF Purpose 9: Apply market research to generate audience insights
\end{itemize}
    \item Other internal uses
\begin{itemize}
        \item TCF Purpose 10: Develop and improve products
    \end{itemize}
\end{itemize}

We consider the last (TCF Purpose 10) to be non-GDPR compliant, as it is vague and ambiguous (what product? to which end?). However, this could be fixed by further specifying purposes. TFC purposes 5 to 8 are also too vague when considering that is the counter-part for the user (details of the service being offered as a result of lending personal data), and thus, key in the consent decision. In this context, our transparency scheme proposal assumes a level of specificity similar to the one used in the following examples:


\begin{itemize}
    \item Route planning
    \item Product recommendation
    \item Language translation
    \item Image generation
    \item Conversational agents
\end{itemize} 

This paper uses a list of purposes aggregating TCF purposes 1, 2 and 4, and adding a list of specific contents and services for purpose 3, as exemplified above. Nonetheless, the proposal of a complete list of purposes is out of the scope of this work.

\section{AI TRANSPARENCY SCHEME}\label{sec:scheme}

An informed society requires the most effort from society itself, both in education (individuals must get familiarised with new concepts) and accountability (individuals are forced to handle added responsibilities). 
The AI community is an expert in the field and can support this effort by devising communication mechanisms which target all of society, tackle all the main issues, and do it with adaptable levels of detail.

With these goals in mind, we define a transparency scheme for systems with AI services. The scheme has two columns and three levels. The columns, cover the two main sources of ethical issues: \textit{Data Privacy} (Figure~\ref{fig:data_privacy}) and \textit{AI Transparency} (Figure~\ref{fig:ai_trans}). The three levels provide an increasing amount of detail.

The first level is the coarsest level of transparency for both columns. It provides the information needed to empower users to a minimal level of autonomy. We use visual signs because these simple representations can convey complex concepts in a quick and non-intrusive manner. This capability is proved daily in domains like mobility, emergency management, law enforcement or user interaction. In fact, icons have previously been consider to illustrate data privacy information~\citep{rossi2020making}. We define a set of visual signs for each column separately.

The second level of transparency must provide more details and more fine-grained decision space. Personalized choices based on multiple preferences must be enabled. For this purpose, we use factsheets, which have been proven effective at gathering informed decisions  \citep{utz2019informed}.
Through factsheets, we summarize the characteristics of each service and enable decisions made on a case-by-case basis.

The third and last level of transparency contains the details available, which should enable a \textit{minimal\/} ethical evaluation for guiding user interaction. Either column contains a different list of public details made available to the user.

\subsection{Data Privacy}\label{sub:dataprivacy}

Many AI systems in use today are data-hungry. Machine learning, the most popular branch of AI, is based on the premise of data. And the more data it can get, the better it may perform. 
At the same time, what generates the most data in the world today is society - billions of people acting and interacting digitally in a globalised world. The popularisation of AI systems feeding on personal data was unavoidable.

While AI provides many valued services, currently people are often unaware (and uninformed) of the fact that an AI system is feeding on their data, and the possible implications with regards to their privacy. A situation which is by itself unethical. 
To advance towards an ethical use of personal data by AI systems, it is necessary to inform users that information may be recorded, stored and further exploited. 
The specifics on which information is to be provided for data privacy has already been legislated upon through GDPR, and will guide the design of the data privacy pillar of the proposed scheme. On the other hand, there is no legislation for AI transparency, the other proposed pillar, which will entail a less contextualized proposal. 

\subsubsection{Visual signs}


As the first level of information, one that is to be always visible, we use a feature that can enable the most basic users' policies and decisions. 
Since we are in the context of a user interacting with an AI system, this feature is (related with) the use of private data within the system, our (or) outside of it. Gathering and using private data generated by a service for improving that service has specific ethical implications. Doing it to feed a different service, has others. 
Thus, for personal data and privacy, we propose three different signs, where one and only one always holds for each user-based AI system: 

\begin{itemize}
    \item \textbf{Personal data not gathered} No active or passive information associated to the user is recorded or disseminated by the system. The system may store information about its own use, but if it does so, it is in an anonymised manner,  not linking interactions with users. In this setting, it is impossible to map together two otherwise independent interactions generated by the same user. This situation is illustrated with a closed lock.
        \begin{minipage}{.74\textwidth}
            \vspace{0.2cm}
           \textit{Example:} A recommender system of an online shop, may store which items are most frequently browsed, when and how. However, it does not store information regarding which items were previously seen or bought by the same user. This behaviour includes not gathering personal details like geographic location, previous activity, access details, etc.
        \end{minipage}
        \begin{minipage}{.18\textwidth}
                   \centering
            \includegraphics[scale=0.3]{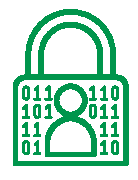}
        \end{minipage}
\end{itemize}

\begin{itemize}
    \item \textbf{Personal data is stored and/or used in this system}: The system may be collecting information from the user. The collected information may only be used by the system in which it was produced, and it is never distributed outside of it. This setting is illustrated with an open lock.
        \begin{minipage}{.74\textwidth}
            \vspace{0.2cm}
           \textit{Example:} An email client learns the writing style of a user by processing her sent emails. This behaviour allows an AI service to learn \textit{how\/} to propose automatic responses, and another one to make grammar suggestions. The user emails, or any derived or learnt representation of these, is never accessed by a different system or third party. In this example, the use of the email data by a different system to provide targeted ads would be forbidden.
        \end{minipage}
        \vspace{0.2cm}
        \begin{minipage}{.18\textwidth}
                   \centering
            \includegraphics[scale=0.18]{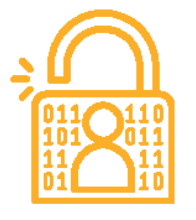}
        \end{minipage}
\end{itemize}

\begin{itemize}
    \item \textbf{Personal data may be stored, exploited, and distributed to third parties}: The system is collecting the user's information. This data could be used for services outside of the system, and also be distributed to third parties. This setting is illustrated with an open lock from which data flows.
        \begin{minipage}{.74\textwidth}
            \vspace{0.2cm}
           \textit{Example:} A video streaming platform stores the history of watched videos by users, linking the data to their profiles. The platform shares this information with a music streaming service, which then uses it for feeding a music recommender system.
        \end{minipage}
        \begin{minipage}{.18\textwidth}
                   \centering
            \includegraphics[scale=0.29]{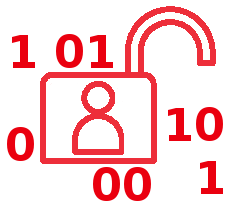}
        \end{minipage}
\end{itemize}

\subsubsection{Data Privacy Factsheet} 

The second level of Data Privacy contains a factsheet, listing all services that request access to personal data. Row by row, services declare what personal data is needed, for what, for how long, and who will be accessing it.  Table \ref{table:factsheet_privacy} provides more details on the information reported by services, together with examples.

\begin{figure}[b]
\centerline{\includegraphics[height=2.7in]{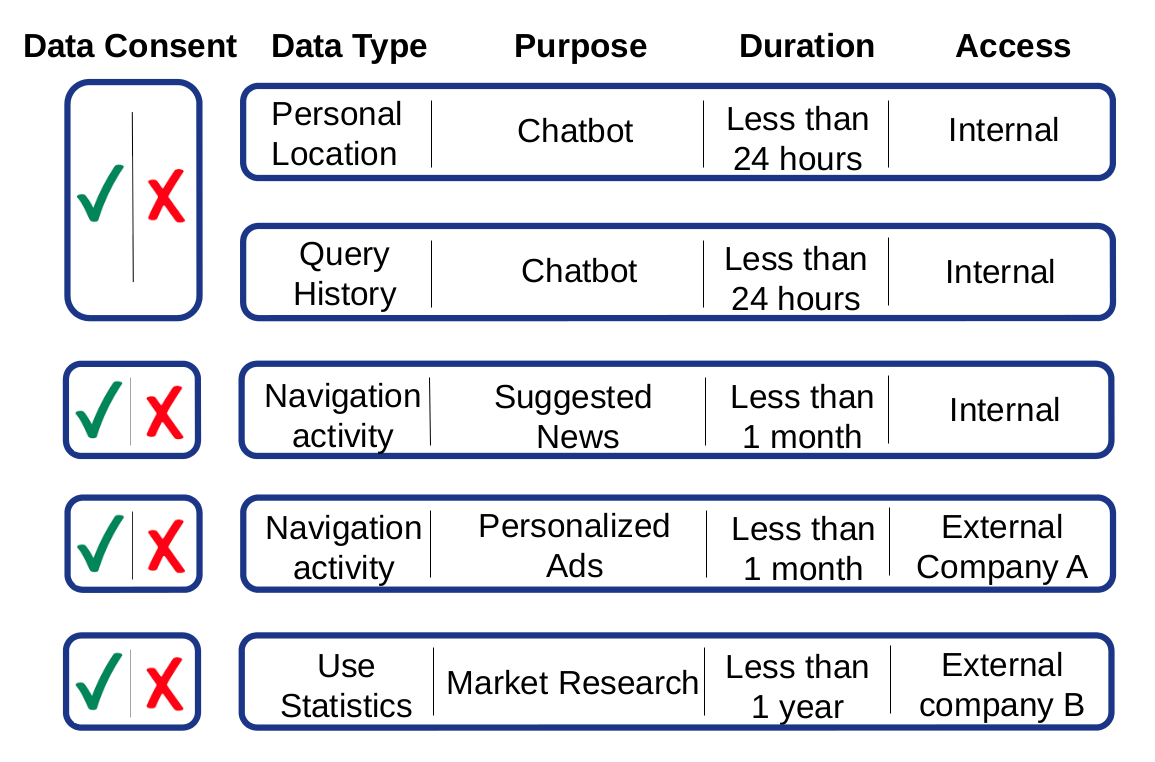}}
\caption{Example of a factsheet for Data Privacy. Each row is a service requesting access to personal data.} 
\label{fig:data_check_sample}
\end{figure}

With this information, users can grant or deny access, taking into account the particularities of each case. To facilitate interaction, with the system, the consents of data services that serve a common and inter-dependable purpose may be aggregated for request/grant. 
An example of a factsheet is shown in Figure \ref{fig:data_check_sample}. This design displays all minimal information and allows users to make quick, service-wise decisions. In compliance with explicit consent, all options should be disabled by default. 

\begin{table*}[t]
\caption{Details provided by services accessing personal data. Guideline for the factsheet at the second level of Data Privacy.}
\begin{tabular}{lll}
\textit{Detail} & \textit{Category} & \textit{Possible values} \\
\hline
Which personal data?& Data types & location, images, navigation,\\
&& use statistics\\
\hline
For which purpose?     &   Purposes      &   personalized ads, language \\
&&translation, market research, \\
&&route planner, \etc \\
\hline
For how long? & Range of periods  &less than a day, less than a\\
&& month, less than a year,\\
&& a year or more \\
\hline
Who has access to it? & Entities & company A, government B, \\
&&conglomerate C \\
\hline
\end{tabular}
\label{table:factsheet_privacy}
\end{table*}

\subsubsection{Full detail: Personal Data}\label{sec:L3_data}


GDPR again inspires the full detail section of the display regarding personal data. In Article 15, this regulation specifies that individuals (also known as data subjects) have the right to access their personal data. 
This is known as the subject or subjective access. This connection would ideally be instantaneous so that people know the status of their privacy at all times. For this purpose, we consider a request form at this level, with additional security measures. 
According to GDPR, services have a maximum of one month to satisfy data requests, and in most cases fees cannot be charged to process them. We consider this time frame should be reduced, eventually reaching the level of seconds.

The request form manages users' access to their personal data, and it contains an option to request all personal data stored. It also contains filters to access data based on the categories of the second level factsheets (data type, purposes, \etc).
Beyond the request form, at this level, we propose a dashboard. Since this is a topic of research on its own right~\citep{bier2016privacyinsight}, only certain charactersitics of it will be discussed here. Particularly, the fact that once personal data is made accessible to its owner, the dashboard must enable them to handle it, as well as their related rights. The functionalities of this dashboard should include:
\begin{itemize}
    \item Browsing the private data stored by the system;
    \item Modifying the restrictions to be applied to the storage and processing (\ie opting in or out of some cases);
    \item Access to details on the source of the data, mainly if this was not obtained directly from the user;
    \item Specifying rectifications on the data;
    \item Demanding the erasure of some or all of personal data stored, and
    \item Issuing a legal complain with a supervisory authority
\end{itemize}

Although we do not propose a specific display for this level here, we anticipate this would have to be a full-screen device.

\subsection{AI Transparency}

AI is typically guided by an algorithm, a model, or both. Algorithms define AI behaviour and are represented through code. Models define the context in which the AI operates, and can be represented by data and/or code. 
If we do not know the algorithm, we cannot know what is the AI doing. If we know the algorithm, but we do not know the model, we can verify the AI behaviour but not its purpose. Any sort of \textit{bias\/} could be encoded within the AI through a model.

For a comprehensive evaluation of an AI system, both code (\ie algorithms and models) and data (\ie models) are needed. This is ideal from a transparency point of view, but it may not always be desirable for the sake of privacy. Sometimes hiding a model provides better security, and sometimes an algorithm is simply private. One way or the other, users must always have the right to know the degree of transparency of every AI system they interact with so that they can take that into account when deciding how to use them, or if using them at all.

A different transparency requirement is related to the results produced by an AI system, instead of the AI system itself. 
Furthermore, that is objectivity or subjectivity of the responses with regards to the user identity. With subjective AI, society must be particularly watchful for traces of manipulative or misguided behaviour, including the promotion or existence of filter bubbles, echo-chambers, partisanship and propaganda\dgnote{Add references to each item in list}. 
Within subjective AI, we include all possible types of personal bias, also those naturally hidden within data and not under human control. For this reason, objective AI can only be guaranteed by forbidding AI algorithms and models to access personal data.

\subsubsection{Visual signs}

For AI transparency, we propose two visual signs. One for code and data transparency and one for objectivity. The first 
includes a representation of both code and data, and their availability. Data is represented through a cylinder (or database) with the text \textit{01} written within. Code is represented by a rectangle (or screen) with the text \textit{$<$Section etminus$>$}. Variants of the sign are generated by changing the colour or the visibility of both components. We introduce three cases that we consider  cover the majority of practical cases:

\begin{itemize}
    \item \textbf{Open AI services}: The source code of all AI algorithms and models running in the system are public to the user. All data used for fitting and deploying these algorithms and models are public to the user. This openness includes code, documentation, data and metadata. This setting is illustrated with a white box of data feeding a white box computer.
    
        \begin{minipage}{.68\textwidth}
            \vspace{0.2cm}
           \textit{Example:} A word embedding model used for suggesting results to queries, has the following data available to the user: Trained model, training code for the model, data used for training the model and description of the data used when running it. This is the only case where the existence of \textit{bias\/} in the system can be audited.
        \end{minipage}
        \begin{minipage}{.25\textwidth}
                   \centering
            \includegraphics[scale=0.40]{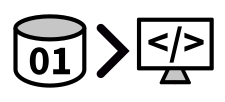}
        \end{minipage}
\end{itemize}

\begin{itemize}
    \item \textbf{Public AI services}: The code of all AI algorithms and models running in the system are public to the user. This includes source code and documentation. However, the data used to fit or run one or more of these algorithms and models is not available to the user. Thus, its behaviour is not fully auditable. This setting is illustrated with a black box of data feeding a white box computer.
    
        \begin{minipage}{.68\textwidth}
            \vspace{0.2cm}
           \textit{Example:} A service for image manipulation fueled by a GAN. The source code of the GAN is available, but the data used to train the model it is not. The service is not open because the existence of bias can not be verified.
        \end{minipage}
        \begin{minipage}{.25\textwidth}
                   \centering
            \includegraphics[scale=0.40]{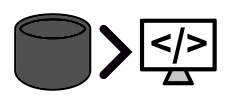}
        \end{minipage}
\end{itemize}

\begin{itemize}
    \item  \textbf{Opaque AI services}: The source code of one or more AI algorithms or models running in the system is not available.  This setting is illustrated with a black box of data feeding a black box computer.
    
        \begin{minipage}{.68\textwidth}
            \vspace{0.2cm}
           \textit{Example:}  A social network suggests (new) contacts and content for users to follow. However, the code and the data powering such service is not available for the users. The motivations and mechanisms driving the recommendations are unknown; the system is opaque.
        \end{minipage}
        \begin{minipage}{.25\textwidth}
                   \centering
            \includegraphics[scale=0.40]{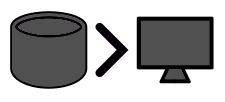}
        \end{minipage}
\end{itemize}

The second visual sign of AI transparency regards the objectivity or subjectivity of the responses provided by AI services. This sign should always be visible together with the previous one. To quickly and clearly convey the idea of objectivity, we propose a pair of complementary signs where only one holds for each AI system:

\begin{itemize}
    \item \textbf{Indistinct information}: The information provided by the system is independent of the identity and behaviour of the user. Any other user producing the same explicit input would receive the same output. User data is inaccessible to the computation.  This setting is illustrated with a round information sign.
    
        \begin{minipage}{.75\textwidth}
            \vspace{0.2cm}
           \textit{Example:}  A search engine produces results to users queries. These results are retrieved using only the query terms, and not using or considering any 
           user-specific information (\eg location, history, device, \etc.
        \end{minipage}
        \begin{minipage}{.18\textwidth}
                   \centering
            \includegraphics[scale=0.55]{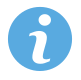}
        \end{minipage}
\end{itemize}

\begin{itemize}
    \item \textbf{Personalised information}: The information provided by the system is or may be subjective with regards to the user identity or behaviour. Any 
    access to personal data by the system enables this scenario. This setting is illustrated with a target sign.
    
        \begin{minipage}{.75\textwidth}
            \vspace{0.2cm}
           \textit{Example:} A video streaming platform offers a variable set of contents to its users. When deciding which content becomes available to each user, the streaming history and profile are taken into account. The system is providing personalised information, and bias may exist.
        \end{minipage}
        \begin{minipage}{.18\textwidth}
                   \centering
            \includegraphics[scale=0.55]{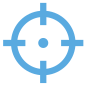}
        \end{minipage}
\end{itemize}

\subsubsection{AI Transparency Factsheet}\label{subsec:ia_2}

\begin{figure}[b]
\centerline{\includegraphics[height=1.40in]{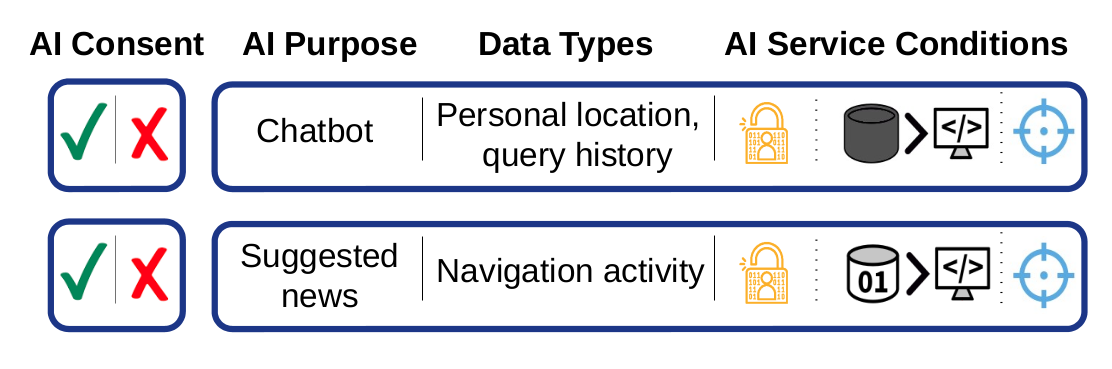}}
\caption{Example of the factsheet for AI Transparency. Each row is an AI service of the system.} 
\label{fig:ai_check_sample}
\end{figure}

The factsheet at the second level of AI transparency contains a list of all AI services running in the system. 
This list may contain entries also found in the Data Privacy factsheet and aggregations of those (a single AI service may request access to several data sources). 
For each AI service, the AI Transparency factsheet specifies its purpose (only one, as specific as possible) and the personal data types it needs for running (none, one or more). It also displays its own specific visual signs with regards to Data Privacy (personal data handling) and AI Transparency (openness of code/data, and objectivity). Figure \ref{fig:ai_check_sample} shows an example of this factsheet.

The combination of all visual signs in the second level factsheet defines the first level system-wide signs for AI Transparency. 
This corresponds to the most restrictive set of each of three signs. \eg a system is subjective if a single service in it is subjective. 
To facilitate navigability, this factsheet may link to several other components of the scheme, like the Data Privacy factsheet, the data request form and the dashboard.
But most prominently, to the full detail section AI Data and Code, detailed next. 

\subsubsection{Full detail: AI Data and Code}


For each AI service in the system with either public code, open data, or both, the full detail level provides an interface to it. 
In the case of source code, it includes minimal documentation as well. In the case of data, it includes anonymized pre-training (if any) and training data, together with metadata describing the nature and structure of the data (\eg source, description of attributes). 
All these full details must be provided for those AI services that are claimed to be open or public in the corresponding second-level factsheet. We expect AI applications on critical domains (\eg healthcare, education) to reach this level of transparency.

\section{VISUAL SIGNS IN ACTION}\label{sec:practice}

The efficiency of a transparency scheme is mostly defined by the way it is implemented and integrated with user interfaces.
If the integration is flawed (\eg intrusive or imperceptible) or it is ill-intentioned (\eg misleading or manipulative), the transparency effort will be undermined \citep{nouwens2020dark}.
To avoid those side-effects, the transparency scheme is to be integrated into the system by design, always visible at some level, and occasionally occupying the main display.
Considering the current trends in AI systems, we advocate for integrating the transparency scheme within browsers (for web applications), operating systems (for client applications) and APIs.
The visual signs proposed were designed to be interpretable even when shown at a small scale (\ie the size of an icon in a browser bar). To illustrate this, some figures in this paper are purposely plotted small.

According to GDPR Article 6 on the lawfulness of processing, the interaction between users and the transparency scheme is to be ruled by \textit{explicit consent}. This is indeed one of the main pillars of personal data rights. 
From the moment a user connects or makes the first request to an AI system, personal data is exposed, and user-AI interaction becomes possible. To inform users of this possibility beforehand so that explicit consent can be granted, the transparency scheme is to be shown on the main display when a user first connects to an AI system. 
Meanwhile, the AI system remains waiting in the background, invisible and unable to access any personal data until explicit consent is given. 

\begin{figure}[ht]
\centerline{\includegraphics[height=0.5in]{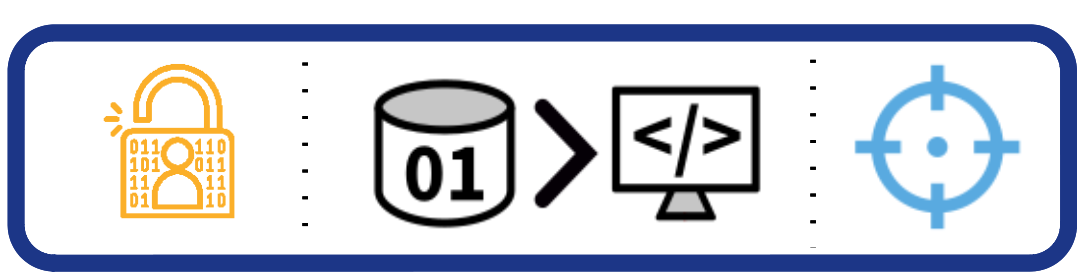}}
\caption{Example of visual display for an AI system which uses personal data to provide targeted content, while having both data and models openly available.} 
\label{example_some}
\end{figure}

As an example, let us consider a web page which includes several AI services. Only one of these services runs on the landing page. 
This AI service is an open-source, rule-based system that adapts the content of the front page to the user location and device. When a user connects to the front page, only the location and no further personal data is gathered. In such a system, when a user connects to the front page for the first time, before any content of the system is shown or any data is gathered, the user is prompted with the icons shown in Figure \ref{example_some}. 

From this view, the user can directly grant or deny consent. In the first case, it will proceed to the front page, location will be gathered, and personalized information will be provided. In the second case (deny), the AI service may provide a version of it without the denied functionalities, or it can reject to serve the user.

To keep users engaged with the transparency scheme, the first level of visual signs is to be always on sight. This is a reminder of the rights granted at that moment, and the applying terms to any ongoing interaction. To minimize interference, the signs should be displayed small (\eg the size of a browser bar button). These signs also provide direct access to the complete transparency scheme at any time.

\begin{figure}[ht]
\centerline{\includegraphics[height=.5in]{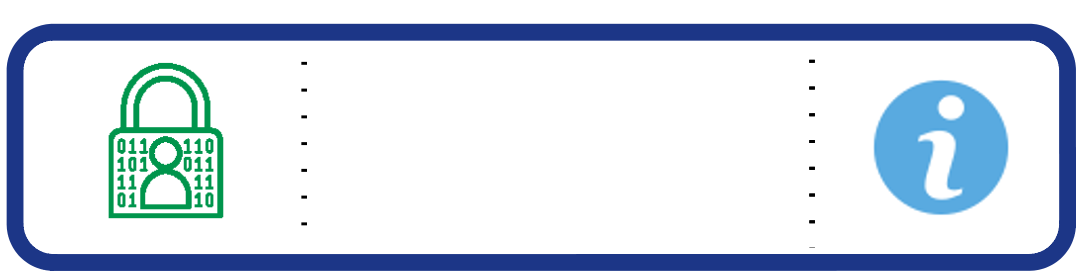}}
\caption{Example of visual display for a system with no personal data gathering or use, containing no AI component, and having an objective behaviour.}
\label{example_none}
\end{figure}

An AI-based system that is not currently running any data gathering or AI services concerning the user may not require explicit consent. Nonetheless, it still needs to be notified. If this is the case, the visual signs shown will be those of Figure \ref{example_none}: No personal data is gathered, and all interactions are indistinct of the user.

If at some point, the conditions change (the system starts accessing personal data or running AI-based services), this should be preemptively notified to the user (following the example of Figure \ref{example_some}). 
This display illustrates the changes in the conditions, and enable the behaviours defined above (grant, deny, inspect \& refine). 
It is important to observe that explicit consent requires that all changes of conditions will be notified: a user who has consented to personalized content for a set of services, is to be notified if a new service with personalized content wants to be activated. Notice however that explicit consent does not need to be given twice for the same service.

In practice, there are some combinations of visual signs that are not possible. The \textit{Indistinct information} icon requires either the \textit{Personal data not gathered} icon, or the \textit{Open AI services} icon. Otherwise the objectivity of the system could not be guaranteed.

\subsection{Visual Signs Today}\label{sec:today}

To assess the current transparency in the context of our proposal, we conduct an analysis on some of the most popular websites worldwide, finding which set of visual signs would apply to them. This study is made based on a review of their data privacy and cookies consent forms conducted between November 2021 and January 2022. When information found is incomplete or not specific enough, the most restrictive policy was applied. Table \ref{tab:today} shows the resulting correspondence.

\begin{table}
    \begin{center}
        \begin{minipage}{324pt}
            \caption{Visual signs corresponding to some of the top worldwide web pages, all of which include AI. Based on a review of their consent forms done during November, 2021.}\label{tab:today}
            \begin{tabular}{c|c}
            \textit{URL} & \textit{Applicable Visual Signs} \\
            www.google.com &  \parbox[c]{6em}{\includegraphics[width=0.7cm]{figs/data_privacy_distributed.png}}
\parbox[c]{3cm}{\includegraphics[width=1.5cm]{figs/ai_private.png}}
\parbox[c]{1cm}{\includegraphics[width=0.7cm]{figs/ai_target.png}}\\
www.facebook.com &  \parbox[c]{6em}{\includegraphics[width=.7cm]{figs/data_privacy_distributed.png}}
\parbox[c]{3cm}{\includegraphics[width=1.5cm]{figs/ai_private.png}}
\parbox[c]{1cm}{\includegraphics[width=.7cm]{figs/ai_target.png}}\\
www.youtube.com &  \parbox[c]{6em}{\includegraphics[width=.7cm]{figs/data_privacy_distributed.png}}
\parbox[c]{3cm}{\includegraphics[width=1.5cm]{figs/ai_private.png}}
\parbox[c]{1cm}{\includegraphics[width=.7cm]{figs/ai_target.png}}\\
www.instagram.com &  \parbox[c]{6em}{\includegraphics[width=.7cm]{figs/data_privacy_distributed.png}}
\parbox[c]{3cm}{\includegraphics[width=1.5cm]{figs/ai_private.png}}
\parbox[c]{1cm}{\includegraphics[width=.7cm]{figs/ai_target.png}}\\
www.twitter.com &  \parbox[c]{6em}{\includegraphics[width=.7cm]{figs/data_privacy_distributed.png}}
\parbox[c]{3cm}{\includegraphics[width=1.5cm]{figs/ai_private.png}}
\parbox[c]{1cm}{\includegraphics[width=.7cm]{figs/ai_target.png}}\\
www.amazon.com &  \parbox[c]{6em}{\includegraphics[width=.7cm]{figs/data_privacy_distributed.png}}
\parbox[c]{3cm}{\includegraphics[width=1.5cm]{figs/ai_private.png}}
\parbox[c]{1cm}{\includegraphics[width=.7cm]{figs/ai_target.png}}\\
www.netflix.com &  \parbox[c]{6em}{\includegraphics[width=.6cm]{figs/data_privacy_distributed.png}}
\parbox[c]{3cm}{\includegraphics[width=1.5cm]{figs/ai_private.png}}
\parbox[c]{1cm}{\includegraphics[width=.7cm]{figs/ai_target.png}}\\
www.reddit.com &  \parbox[c]{6em}{\includegraphics[width=.6cm]{figs/data_privacy_distributed.png}}
\parbox[c]{3cm}{\includegraphics[width=1.5cm]{figs/ai_private.png}}
\parbox[c]{1cm}{\includegraphics[width=.7cm]{figs/ai_target.png}}\\
www.tiktok.com &  \parbox[c]{6em}{\includegraphics[width=.6cm]{figs/data_privacy_distributed.png}}
\parbox[c]{3cm}{\includegraphics[width=1.5cm]{figs/ai_private.png}}
\parbox[c]{1cm}{\includegraphics[width=.7cm]{figs/ai_target.png}}\\
www.discord.com &  \parbox[c]{6em}{\includegraphics[width=.6cm]{figs/data_privacy_distributed.png}}
\parbox[c]{3cm}{\includegraphics[width=1.5cm]{figs/ai_private.png}}
\parbox[c]{1cm}{\includegraphics[width=.7cm]{figs/ai_target.png}}\\
www.twitch.tv &  \parbox[c]{6em}{\includegraphics[width=.6cm]{figs/data_privacy_distributed.png}}
\parbox[c]{3cm}{\includegraphics[width=1.5cm]{figs/ai_private.png}}
\parbox[c]{1cm}{\includegraphics[width=.7cm]{figs/ai_target.png}}\\
www.fandom.com &  \parbox[c]{6em}{\includegraphics[width=.6cm]{figs/data_privacy_distributed.png}}
\parbox[c]{3cm}{\includegraphics[width=1.5cm]{figs/ai_private.png}}
\parbox[c]{1cm}{\includegraphics[width=.7cm]{figs/ai_target.png}}\\
www.accuweather.com &  \parbox[c]{6em}{\includegraphics[width=.6cm]{figs/data_privacy_distributed.png}}
\parbox[c]{3cm}{\includegraphics[width=1.5cm]{figs/ai_private.png}}
\parbox[c]{1cm}{\includegraphics[width=.7cm]{figs/ai_target.png}}\\
www.whatsapp.com &  \parbox[c]{6em}{\includegraphics[width=.6cm]{figs/data_privacy_open.png}}
\parbox[c]{3cm}{\includegraphics[width=1.5cm]{figs/ai_private.png}}
\parbox[c]{1cm}{\includegraphics[width=.7cm]{figs/ai_target.png}}\\
www.wikipedia.org &  \parbox[c]{6em}{\includegraphics[width=.5cm]{figs/data_privacy_locked.png}}
\parbox[c]{3cm}{\includegraphics[width=1.5cm]{figs/ai_private.png}}
\parbox[c]{1cm}{\includegraphics[width=.7cm]{figs/ai_nobias.png}}\\
www.duckduckgo.com &  \parbox[c]{6em}{\includegraphics[width=.5cm]{figs/data_privacy_locked.png}}
\parbox[c]{3cm}{\includegraphics[width=1.5cm]{figs/ai_private.png}}
\parbox[c]{1cm}{\includegraphics[width=.7cm]{figs/ai_nobias.png}}\\
www.github.com &  \parbox[c]{6em}{\includegraphics[width=.5cm]{figs/data_privacy_locked.png}}
\parbox[c]{3cm}{\includegraphics[width=1.5cm]{figs/ai_private.png}}
\parbox[c]{1cm}{\includegraphics[width=.7cm]{figs/ai_nobias.png}}\\
www.wordpress.org &  \parbox[c]{6em}{\includegraphics[width=.5cm]{figs/data_privacy_locked.png}}
\parbox[c]{3cm}{\includegraphics[width=1.5cm]{figs/ai_private.png}}
\parbox[c]{1cm}{\includegraphics[width=.7cm]{figs/ai_nobias.png}}\\
www.openstreetmap.org &  \parbox[c]{6em}{\includegraphics[width=.5cm]{figs/data_privacy_locked.png}}
\parbox[c]{3cm}{\includegraphics[width=1.5cm]{figs/ai_private.png}}
\parbox[c]{1cm}{\includegraphics[width=.7cm]{figs/ai_nobias.png}}\\
\hline
            \hline
            \end{tabular}
        \end{minipage}
    \end{center}
\end{table}

Current legislation on data privacy (\eg GDPR) has increased transparency on personal data. As shown in Table \ref{tab:today}, while the majority of services gather and disclose personal information, others are taking a more restrictive approach, maintaining a complete anonymity and privacy of users (first column of signs). This is most common in sites related with the open source movement, who also provide objective access to some of their functionalities (third column). There is one case, \textit{whatsapp} were conditions did not specify the sharing of personal data with third parties. This is most likely caused by the lack of ads in their website, which is responsible for the majority of data privacy transactions.

On the other hand, the lack of regulation regarding transparency in AI methods has resulted in generalized opacity regarding AI-based services. This is shown in the second column of Table \ref{tab:today}, where all reviewed sites get assigned the least transparent labeling. Even open source sites, which typically open their frontend but not their backend (where AI services run) fall short in this category. Let us remark that some sites, like \textit{wikipedia}, provide the models and data used to train \textit{some} of their AI services in an exercise of transparency. However, since this is not complete (some AI services remain under-specified) these sites still get assigned the opaque AI services icon. Notice we assigned the AI opaque icon to all services, even when we could not confirm their use of AI tech. Preventively, this was assumed by default.

\subsection{Domains of Application}\label{sec:domains}

The proposed visual scheme is compatible with AI services running in a wide variety of domains. Each domain must implement certain interaction mechanisms to guarantee the transparency principles, although how are these specifically implemented can vary significantly, as we will see next. The three main interactions we consider are:


\begin{itemize}
    \item Granting/revoking data consents: whenever applicable, the user should always have easy access to update data consents.
    \item Requesting models or code: whenever the AI system claims to have an open model or code, the user should have easy access to instantly download them.
    \item Requesting auditing traces: at all times, the user should have easy access to instantly download the full trace of the use of their data inside the AI system, including data being shared to third parties.
\end{itemize}

Depending on the specific medium with which the user is using the AI system, we can distinguish several possible channels for interaction.

\subsubsection{Frontend UIs}

These include mainly web and desktop applications. One possible mode of interaction is the use of banners floating at the top or the bottom of the UI (user interface). The main advantage for using banners is that users are already used to reading and interacting with this kind of visual component, given their application to cookie management.

We recommend that the UI is designed in a way such that this type of component is fully integrated into the interface, especially for the visual signs which are designed to be compact. For example, the component could be set in a specific place common to all or most views of the application such as the top right corner. On one hand, this would state that the intentions of the owners of the AI system are to be transparent about the usage of the user's data. On the other hand, this would serve as a permanent reminder for the user about this usage.

Transparency, by definition, has to apply to all users regardless of age, cultural background or physical or psychological conditions. We propose combining the design requirements we propose with those applicable requirements specified in Universal Design \citep{steinfeld2012universal,johnson2017designing} that may apply. For example, accounting for blindness or color blindness by enabling speech-to-text or allowing a change in font size.

\subsubsection{Mobile apps}

Mobile apps can also use the same types of UI components as web or desktop applications. However, mobile devices allow for added communication channels that could contribute in providing transparency: push notifications and OS-controlled capabilities.

Push notifications, due to their nature, would not be useful for enabling a permanent reminder, but can be used as a complement to the aforementioned methods. Specifically, the user could voluntarily subscribe to triggers related to the use of their data, such as when a model has been refreshed using recent user's data, or when a particular service has started using it.

OS-controlled capabilities are a mechanism provided by the operating systems (\eg Android, iOS) that allow users to have strict control over important features of their devices. In most cases this control is managed by explicitly granting or revoking access to the device resources (\eg camera, microphone, GPS location) for specific apps or by having the operating system proactively asking the user based on the capabilities defined in the app manifest.

Unfortunately, there is still no transparency capability enabled in these operating systems. This is a topic of ongoing discussion, because granting access to a resource for an app is sometimes too coarse (for example, the app could be using the resource for purposes not explicitly stated, even in the background when the app is not being used). We believe that improvements could be done in operating systems in this direction. In any case and while that is not the case, we propose the definition of settings, as similar as possible to the ones already existing for granting/revoking access to device resource, for the management of data consents.

\subsubsection{APIs}

While APIs are not formally a type of user interface, they are still a form of communication that could involve the user, by means of a proxy. For example, an advanced user can use the Twitter API to gather data about themselves or act on their behalf.

APIs are usually documented, not only in the form of textual documentation but also inside the requests and responses using machine processable patterns. For example, if the response includes an HTTP header with a key similar to \emph{X-Rate-Limit} we will know, at the very least, that the server will limit the allowed frequency of requests. This header could also include the specific amount, or if the rate limit has been exceeded.

We propose the use of headers for those APIs that allow a machine to interact with an AI system in behalf of a user. This headers should implement the scheme described in Sections \ref{sec:scheme} and \ref{sec:practice}. For example:

\begin{itemize}
    \item \emph{X-Personal-Data} = [not gathered, may be stored, may be exploited]
    \item \emph{X-Transparency-Code-Data} = [open, public, opaque]
    \item \emph{X-Transparency-Objectivity} = [indistinct, personalised]
\end{itemize}

\section{TRANSPARENCY BY DESIGN}\label{sec:integration}

The scheme proposed in this work (detailed in Section \ref{sec:scheme}) requires a system capable of 1) presenting correct and truthful information to the user in terms of factsheets, and 2) enforcing the preferences explicitly set by users, including options for auditing the use of their data by the system. Both of these requirements are platform and software agnostic, and can be implemented on different domains like the ones specified in Section \ref{sec:practice}. 

For broadening the scope of our work, and facilitate adoption, let us introduce a domain independent software design paradigm to guarantee the transparency requirements of the proposed scheme. What we called, transparency by design. Previous contributions tackle the same issue, from a social and legal perspective~\citep{rossi2020transparency}. Here we concert ourselves with the technical one. We define high level, flexible guidelines for integration, mimicking what can be found in other parallel approaches for tackling pervasive cross-functional requirements such as \emph{security by design} and \emph{privacy by design} \citep{gurses2011engineering}.



\begin{figure}
\centerline{\includegraphics[width=4in]{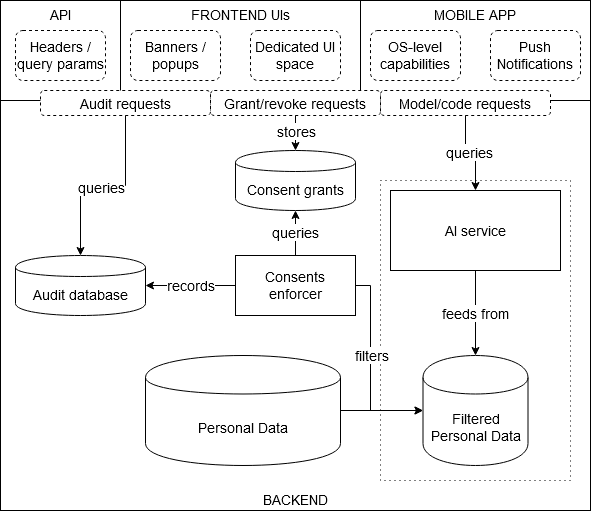}}
\caption{Relevant elements for integration of visual signs, for both factsheets visualisation and enforcement of user policies} 
\label{fig:integration}
\end{figure}

Our proposal is summarised in Figure~\ref{fig:integration} and includes a set of general recommendations that can be implemented as specific components or methodological guidelines on a wide variety of domains. The goal is to enable the system to be designed end to end in a manner that automatically enables transparency for the final user.

\dgnote{Maybe some introductory text to the figure? What am I looking at? Color/shape codes? what is the dashed box around AI service?}


In order to properly integrate privacy and transparency, it is not enough with being explicit in the purposes and the specific data consents that the user must know or manage. It is still necessary, and equally as important, to design the AI system in a way that enforcement of such consents and the capability of auditing the use of the data are both guaranteed.

Figure~\ref{fig:integration} includes an AI service (which could be one or many, and which could be owned by the AI system or by a third party) and storage for personal data from users. Around these, to complement the, we integrate the following high-level concepts (or components if the mapping fits the design): a consent grants database, a consent enforcer, and an audit database.\dgnote{And a filtered personal data database?} The following subsections describe their corresponding purposes.

\subsection{Consent Grants Database}

For the sake of security, accessibility and accountability, AI systems should include some form of storage for all the consents explicitly granted by users. By default all consents are denied. Thus, only the consents which have been explicitly granted should be stored. If a consent was granted and later revoked, it should be removed from the database. However, periodical snapshots of the database should be stored for a limited period of time, so that auditing is possible.

\subsection{Consent Enforcer}

The AI services should never use personal data directly. Because the permission of using the data will depend on the specific service or purpose combined with the data consents declared by the user, the AI system has to include, in its design, a component processing the data consents and dynamically generating a filtered version of the user personal data that complies with such consents. We call this component the consent enforcer, which should work as a hard filter and be as simple as possible.

\subsection{Filtered Personal Data}

The outcome of the consent enforcer is the filtered personal data, available for a specific AI service. This filtered dataset should serve two purposes. On one hand, it is the only personal data AI services should be able to process. On the other hand, this dataset should be cached or materialised as a database view, in a way that it is accessible and inspectable by users. That is, as described in the third level of data Privacy (see \ref{sec:L3_data}).


\subsection{Audit Database}

In order to ensure transparency, the usage of data must be auditable by the owner of the data. This means that whoever operates with that data has to be always ready to disclose that usage and therefore have the capacity to be held accountable \citep{guts2009reliable}.

A first step for enabling this is to record every single instance of usage of the data. As every usage is, in our proposal, already authorised or denied by the \emph{Consent Enforcer}, it should be enough with recording such operations in a storage that we call \emph{Audit Database}. Each record stored in this component should include, at least:

\begin{itemize}
    \item The origin or trigger of the operation (such as the execution of an AI service, or a consent granting action from a user),
    \item The timestamp,
    \item The specific version of the personal data or any alternative way to identify the contents of this data at any arbitrary point in time,
    \item A formalised description of the filter applied to this data, generated by the consent enforcer
    \item The specific version of the AI service, such as a release version number or a version control commit hash.
\end{itemize}

Although storing this information along with the data versioning may imply considerable overheads, we consider this list to be the minimum set necessary for ensuring transparency as defined in \ref{sec:scheme}.

\section{AI Act Contextualization}

The European Commission has recently released what is known as the \textit{AI Act}~\citep{AIACT}, a set of directives which will guide the development of future AI regulation. While this was released after the contributions of this paper were finished, we cannot ignore its relevance for our work. In this section we review the AI Act from the perspective of the transparency and consent framework proposed here.

In Article 52, the AI Act specifies the obligation to notify the presence of AI when this is \textit{"interacting with humans"}. While this is rather ambiguous, it goes in the direction of our work, which assumes users will always be aware of the presence of AI in the services they consume.

Within the AI Act, AI is categorized in levels of risk based on the sort of application they perform (Article 6 \& 7). We find this incomplete, in the sense that it does not protect users from AI with access to highly sensitive personal information (\eg gender, racial origin, ideology \etc), when the purpose of this AI is not considered critical (\eg selecting news for the reader). Furthermore, while purpose is fuzzy, hard to inspect, specify and enforce (as exemplified by the current implementations of GDPR), data access is much more comprehensible and straight-forward. 

Given the differences in regulation between high-risk AI and the rest, our analysis of compatibility with the proposed transparency scheme will depend on the nature of the AI running in the background (which is typically opaque and unverifiable, as seen in Table \ref{tab:today}). Let us first consider the case of high-risk AI. In this domain, and according to Article 11, 18 and Annex IV, AI services must be transparent enough as to assess if the services provided are objective or subjective. At the same time this also guarantees that the data and code transparency icons will be enforceable.

For the rest of AI services (\ie those not considered as high-risk in the AI Act) there is no requirement for documentation, transparency, or any other sort of user notification. Only a couple of exceptions are considered. Most significantly, the notification of the presence of AI (as stated above), and the labeling of synthetic media generated by AI systems (what is known as \textit{deep fakes}). This puts consumers at a disadvantage, and makes unfeasible the implementation of the work here presented.

\section{PREVIOUS WORK}\label{sec:previous}

In this work, we propose a transparency scheme to empower users on their interactions with AI systems. In the past, there have been some similar contributions, but most of these works precede the publication and enforcement of GDPR, and some of them are not GDPR compliant (\eg opt-out by default). Another significant limitation found in these proposals is their lack of an adaptable level of detail \citep{schaub2015design}. Given the variability in expertise and concern within society, it is essential to provide several levels of specificity to satisfy all social demands of transparency.

In \citep{kelley2009nutrition} authors propose a display based on the nutrition labels used by the food industry. Their final proposal defines a matrix, with rows being types of information, and columns being purposes and entities having access to the data. Each cell has several possible values, such as requiring opt-in, requiring opt-out, \etc. A more recent contribution was presented in \citep{schaub2015design}. Again, restricted to the context of data privacy, authors consider the different aspects that must rule an effective communication mechanism. These include \textit{notice complexity} (which we tackle through three levels of increasing detail), \textit{lack of choices} (which we implement on all three levels), \textit{notice fatigue} (which we reduce through the first, browser-integrated level) and \textit{decoupled notices} (which we solve through a unified scheme). Although the proposal of Schaub \etal has a broader spectrum than ours (\eg it considers privacy in wearables), for those applicable aspects it is well aligned with our own.

\section{CONCLUSIONS}\label{sec:conc}

Ethical principles take time to be put in practice. They can only be enforced through social awareness and consensus, as their principles slowly sink in individuals. An example of such process is the Universal Declaration of Human Rights from 1948. This set of ethical principles was a response to the devastating effect of new disruptive technology on the world (automatic, chemical and nuclear weapons, missiles, aerial warfare, telecommunications, \textit{etc}). Certain uses of this technology enabled an unprecedented capacity for violence that was deemed universally unacceptable. However, as any ethical principles, the Universal Declaration of Human Rights could not be assumed and practiced overnight. It took 30 years, in the late 70s, for humankind to start embracing it. This shows in the volume of prosecution on human right violations, which started to grow in that period, and continued to do so consistently afterwards~\citep{dancy2019behind}. We can expect ethical AI to follow a similar progression until generalized adherence. Our goal is to reduce the adoption time by increasing the awareness and empowering of society. In this paper we propose to do so through transparency, as a tool against the unethical and abusive use of AI technology. With that purpose in mind we define a transparency scheme that covers the essential aspects of AI at different levels of detail.

The scheme proposed in this paper exists in two different contexts. On the side of data privacy, it is a follow up work on legislation already in place. In this case, the contribution is designed to maximize the impact and effectiveness of certain aspects of that legislation. The other context of this work regards AI transparency. In this domain there is no legislation, and current behaviors are unbounded and mostly secretive. Thus, the proposal here is innovative, subject to potential improvement by the community, and requiring legislative enforcement for its widespread adoption.

Although transparency does not prevent unethical behavior on its own, it makes it punishable. In today's situation with no transparency, providers of AI services have no motivation to constraint the reach and secrecy of their technology. Users remain unaware and uninformed, rendering them powerless. With the implementation of the transparency scheme proposed here, users become empowered, enabled to refuse the use of certain services under certain conditions. This alone will shift the balance towards more transparent AI services, seeking the reward of user confidence, and avoiding their reticence towards dark systems.

One key contribution of this paper are the visual signs. Conveying basic features of AI systems visually is necessary for an efficient and effective interaction with all sorts of users. The signs we propose have been designed to guarantee its proper interpretation even if they are shown very small. Beyond sings, factsheets generate a more detailed decision space, enabling more complex decision making processes. Finally, the full detail level provides access to all claimable information. 

To complement the theoretical exercise, we propose a set of \textit{transparency by design} guidelines, which can be used as a general scheme for the integration of AI services. We further discuss implementation mechanisms on several domains, illustrating how feasible it is to put the proposed mechanism in practice in a relatively short time. 

\subsection{Future Work}

The integration of AI in society has a long path to tread. In coherence, this is not a closed proposal, but one open to improvements and updates. We expect this process to span for years, as society and AI technology keep advancing and adapting to one another. The following list includes issues we think are a priority to tackle for the practical use and improvement of the proposed transparency scheme.

\begin{itemize}
    \item GDPR’s subjective access specifies the right to access your personal data under certain conditions. Industry is implementing this right through request forms, the response to which can take up to a month. This is clearly sub-optimal in terms of transparency, and it would be best if users could directly and instantly have access to their stored personal data. Demonstrating that this is technically possible would be a great contribution for transparency.
    \item Factsheets have been shown to produce more informed decisions than a single binary choice \citep{utz2019informed}, however they can still be extended. An interesting future work is to complement these with other methodologies, such as the two-dimensional table displays used in some of the related works, decision trees, or white and black lists. 
    \item For the context of this work, we use the definition of AI service provided  in  Section \ref{sec:terms}.  However, this definition is not specific enough in some key aspects (noticeably, in the "Who has access" factsheet of Data Privacy). The boundaries of data use within large-scale, integrated systems that provide many different services among different platforms remain fuzzy. Specifying these boundaries in the current industrial context represents a huge challenge ahead.  
    \item The definition of purpose in the context of consent is essential for transparency. Properly specified purposes will engage users properly. In Section \ref{subsec:purpose} we briefly discuss the topic, and propose a list of purposes, abstracting some of TCF 2.0 purposes and expanding others. While this is enough to illustrate the level of detail that would be needed for the transparency scheme to work, the topic requires further analysis. Doing a complete and functional list of purposes is another important future work that remains. Finding ways to keep such list continuously updated will also be a challenge.
In this work, we propose a transparency scheme to empower users on their interactions with AI systems. In the past, there have been some similar contributions, but most of these works precede the publication and enforcement of GDPR, and some of them are not GDPR compliant (\eg opt-out by default). Another significant limitation found in these proposals is their lack of an adaptable level of detail  variants in their daily interactions. This could be achieved through a browser extension, as done in the Study \#2 of \citep{nouwens2020dark}. 
\end{itemize}



\bibliographystyle{apacite}
\bibliography{signs}
\end{document}